\documentclass[aps,prd,twocolumn,showpacs,nofootinbib]{revtex4}

\usepackage{epsfig,amssymb,euscript}
\usepackage{amsmath,amscd}

\newcommand{\be}{\begin{eqnarray}}
\newcommand{\ee}{\end{eqnarray}}
\newcommand{\bea}{\begin{eqnarray}}
\newcommand{\eea}{\end{eqnarray}}
\newcommand{\ba}{\begin{array}}
\newcommand{\ea}{\end{array}}

\newcommand{\nn}{\nonumber \\}

\newcommand{\p}[1]{(\ref{#1})}

\newcommand{\bR}{\mathbb{R}}

\newcommand{\bchi}{{\mbox{\boldmath $\chi$}}}
\newcommand{\bomega}{{\mbox{\boldmath $\omega$}}}

\begin{document}

\title{Concentric Black Rings}

\author{Jerome P. Gauntlett}

\affiliation{Perimeter Institute for Theoretical
Physics\\ Waterloo, ON, N2J 2W9, Canada
\\
\\ (On leave from: Blackett Laboratory, Imperial
  College, London, SW7 2AZ, U.K.)}

\author{Jan B. Gutowski}

\affiliation{Mathematical Institute, Oxford University,
\\Oxford, OX1 3LB, U.K.}

\date{\today}

\begin{abstract}
\noindent We present new supersymmetric solutions of five-dimensional
minimal supergravity that describe concentric black rings
with an optional black hole at the common centre.
Configurations of two black rings are found which have the same conserved
charges as a single rotating black hole; these black rings can have
a total horizon area less than, equal to, or greater than
the black hole with
the same charges. A numerical investigation of
these particular black ring solutions suggests that they
do not have closed timelike curves.
\end{abstract}


\maketitle

\section{Introduction}
A remarkable feature of asymptotically
flat black holes in five spacetime dimensions, in contrast to four
dimensions, is that they can have event horizons with non-spherical
topology.  The first example of such a black hole was the discovery
of the ``black
ring'' solution of vacuum Einstein equations by Emparan and Reall
\cite{ER}. This is a rotating black hole with an event horizon of
topology $S^1\times S^2$; the rotation being required to prevent the
ring from collapsing. The black ring of \cite{ER} carries a single
rotation parameter of a possible two in five dimensions. This
solution was further generalised in \cite{HE,EE,RE}.

An interesting recent development is the discovery of a supersymmetric
black ring solution of five-dimensional minimal supergravity
\cite{eemr}.  The solution depends on three parameters which are
uniquely specified by the mass and two independent rotation
parameters, $J_1$ and $J_2$. The black ring also carries electric charge
which is proportional to the mass, as dictated by supersymmetry.  It was
shown in \cite{eemr} that the infinite radius limit of this solution leads
to a supersymmetric straight string solution that was found earlier in
\cite{bena}.  It was also shown in \cite{eemr} that upon setting one
of the parameters to zero, one recovers the rotating black hole solution
discovered in \cite{Tseytlin:1996as,bmpv} (see \cite{gauntlett:99}
for a discussion in minimal supergravity).
Recall that these black holes have an event horizon with
topology $S^3$ and carry a single rotation parameter, $J_1=J_2$.

Since the black ring solution of \cite{eemr} preserves supersymmetry,
one suspects that more general multi-black ring solutions might exist.
The supersymmetric black ring solution of \cite{eemr} was found using
the results of \cite{gghpr} which provided a classification of all
supersymmetric bosonic solutions of minimal supergravity in
five-dimensions.  As noted in \cite{eemr}, the BPS equations solved by
the black ring appear to be non-linear and this obscures the
construction of multiple ring solutions via simple superpositions.
Here we will show, also using the results of \cite{gghpr},
that there is in fact a linear structure underlying the solutions
and that this allows us to construct supersymmetric multi-ring solutions
in a straightforward way. In fact we find that it is possible to
superpose arbitrary numbers of concentric black rings,
and moreover, it is also possible to place a
black hole at the common centre.
The most general solutions have $R\times U(1)$ isometry with the
$S^1$ direction of the Killing horizons of all
of the rings lying on the orbits of the $U(1)$ Killing vector field.

We will show that there are configurations of two black rings
that have exactly the same conserved charges as
the rotating black hole solution of
\cite{Tseytlin:1996as,bmpv}; this cannot
be achieved with a single black ring since it has $J_1 \neq J_2$.
We find that these two black ring configurations can have
a total horizon area less than, equal to, or greater
than the black hole with the same charges. In particular,
and surprisingly, the rings can be entropically
preferred. Moreover, a preliminary
 numerical investigation of these solutions suggests
that these black rings do not have closed timelike curves.

Building on the original work of \cite{stva}, a
set of string microstates was identified in
\cite{bmpv} which account for the rotating black hole entropy.
It would be very interesting to identify the microstates corresponding
to the single black ring. Presumably the entropy of the multi-black rings
would then be associated with the relevant multi-string states.
However, it is interesting that the conserved charges of the black holes
and black rings in themselves cannot be sufficient to distinguish between
different string microstates. For
some discussion of black hole and black string microstate
counting in a related context see \cite{EE,RE,benakraus}.

The paper also includes a simple construction of multi-black strings,
generalising the single
black string solution of \cite{bena} and we briefly
mention some other generalisations of the black ring and black strings.

\section{Background Formalism}
The bosonic sector of
five-dimensional minimal supergravity is Einstein-Maxwell theory with
a Chern-Simons term \cite{cremmer}.
A classification of the most general kinds of
supersymmetric bosonic solutions of this theory was carried out in
\cite{gghpr}. What is relevant here is the case when there exists,
locally, a timelike Killing-vector $V$ that can be constructed as a
bilinear of the preserved supersymmetry parameter.  For this case the
line element can be written as
\be
ds^2 = -f^2(dt+\omega)^2 + f^{-1}
ds^2(M_4) \,, \label{metric}
\ee
where $V=\partial_t$,
$M_4$ is an arbitrary hyper-K\"ahler space, and $f$ and $\omega$ are a
scalar and a one-form on $M_4$, respectively, which must satisfy
\be
d G^+ =0 \, , \qquad \Delta f^{-1} = \frac{4}{9} (G^+)^2 \,,
\label{eom}
\ee
where $G^+\equiv \frac{1}{2}f(d\omega+ *d\omega)$, with $*$
the Hodge dual on $M_4$ and $\Delta$ is the Laplacian on $M_4$. The
two-form field strength is given by
\be
F=\frac{\sqrt
3}{2}d[f(dt+\omega)]-\frac{1}{\sqrt 3}G^+ \ .
\ee
Note that we are using
the conventions of \cite{gghpr} but with a change in the signature of
the metric.

We will only be interested in the case when $M_4$ is flat space,
$\bR^4$, and it will be useful to use the following coordinates
\bea
\label{ghflat}
ds^2(\bR^4)&=&H[dx^i dx^i]+H^{-1}(d\psi+\chi_i
dx^i)^2\nn &=&H[dr^2+r^2(d\theta^2+\sin^2(\theta)d\phi^2)]
\nn
&+&H^{-1}(d\psi+\cos\theta d\phi)^2
\eea
with $H=1/|{\bf x}|\equiv 1/r$ and we observe
that $\chi_i dx^i\equiv \cos\theta
d\phi$ satisfies $\nabla\times {\bchi}=\nabla H$.  The range of the
angular coordinates are $0<\theta<\pi$, $0<\phi<2 \pi$ and
$0<\psi<4\pi$.  This displays $\bR^4$ as a special example of a
``Gibbons-Hawking'' hyper-K\"ahler metric \cite{GibH}, and in
particular we note that $H$ is a single-centre harmonic function on
the base $\bR^3$ with coordinates ${\bf x}$.  We will further demand
that the tri-holomorphic vector field $\partial_\psi$ is a
Killing vector of the five-dimensional metric. The significance of
this is that the most general solution is now specified by three further
harmonic functions, $K$, $L$ and $M$ on $\bR^3$ \cite{gghpr}. In
particular the general solution has \be f^{-1}=H^{-1}K^2+L \ee
and if we write
\be
\omega=\omega_\psi(d\psi+\cos\theta d\phi) +\hat\omega_i dx^i
\ee
then
\be
\omega_\psi=H^{-2}K^3+\frac{3}{2}H^{-1}KL+M
\ee
and
\be
\label{hatomeq} \nabla\times \hat\bomega=H\nabla M-M\nabla
H+\frac{3}{2}(K\nabla L-L\nabla K) \ .
\ee
We also record the relation:
\be
i_{\partial_\psi}G^+=-\frac{3}{2}d(KH^{-1})
\ee
which is useful in determining the harmonic functions given
a solution not written in Gibbons-Hawking form.
Finally we note here that a constant
term in the harmonic function $M$ can always be removed, locally, by a
coordinate transformation that shifts $t$.

\section{The single black-ring solution}
We now recall the solution for a single supersymmetric
black-ring solution presented in \cite{eemr}.
We will rewrite this solution using the
Gibbons-Hawking type coordinates as in \p{ghflat} and extract the
harmonic functions $K,L$ and $M$, which will enable us to then generalise
the solution.

The solution in \cite{eemr} was presented in coordinates in which the
metric on $\bR^4$ is given by
\bea
ds^2(\bR^4) &=& \frac{R^2}{(x-y)^2}
\Big[ \frac{dy^2}{y^2-1} + (y^2-1)d\phi_2^2
\nn
&+&\frac{dx^2}{1-x^2}+(1-x^2)d\phi_1^2 \Big] \,.
\label{base}
\eea
Note that in \cite{eemr} $\phi_1$ and $\phi_2$ were labelled
$\phi$ and $\psi$ respectively.  The coordinates have ranges $-1\leq
x\leq 1$ and $-\infty<y\leq -1$, and $\phi_i$ have period $2\pi$.
Asymptotic infinity lies at $x\to y\to -1$. Note that the apparent
singularities at $y=-1$ and $x = \pm 1$ are merely coordinate
singularities. The orientation is $\epsilon_{y\phi_2 x \phi_1} \equiv
+R^4/(x-y)^4$.  After the coordinate transformation \cite{eemr}
\be
\label{eqn:flatcoords}
\rho \sin \Theta = \frac{R \sqrt{y^2-1}}{x-y}, \qquad \rho \cos \Theta
= \frac{R \sqrt{1-x^2}}{x-y}
\ee
we get
\be
ds^2(\bR^4)=d\rho^2+\rho^2(d\Theta^2+\cos^2\Theta
d\phi_1^2+\sin^2\Theta d\phi_2^2)
\ee
and the additional coordinate
transformation
\bea
\phi_1&=&\frac{1}{2}(\psi+\phi),\qquad
\phi_2=\frac{1}{2}(\psi-\phi)
\nn \Theta&=&\frac{1}{2}\theta,\qquad\qquad
\rho=2 {\sqrt r}
\eea
brings the metric into the Gibbons-Hawking form \p{ghflat}.

The single black ring solution of \cite{eemr} has \be
f^{-1}=1+\frac{Q-q^2}{2R^2}(x-y)-\frac{q^2}{4 R^2}(x^2-y^2)
\label{simplef}
\ee and $\omega= \omega_{\phi_1}(x,y) d\phi_1+\omega_{\phi_2}(x,y)
d\phi_2$, with
\bea
\label{omegas}
\omega_{\phi_1} &=& -\frac{q}{8R^2} (1-x^2) \left[3Q - q^2 ( 3+x+y)
 \right]\,, \\ \omega_{\phi_2} &=& \frac{3}{2} q(1+y) + \frac{q}{8R^2}
 (1-y^2) \left[3Q - q^2 (3 +x+y) \right] \,.\nonumber
\eea $Q$ and
 $q$ are positive constants, proportional to the net charge and to the
 local dipole charge of the ring, respectively. We assume $Q\geq q^2$
 so that $f^{-1} \geq 0$.
It is straightforward to see that
\be
G^+=\frac{3q}{4}(dx\wedge d\phi_1+dy\wedge d\phi_2) \ .
\ee

By writing  $f$ and $\omega$ in the Gibbons-Hawking coordinates
$(r,\theta,\phi,\psi)$ we can determine
the three harmonic functions $K,L,M$. We find that they can all
be expressed in terms of a single harmonic function $h_1$ given by
\be
h_1=\frac{1}{|{\bf x}-{\bf x_1}|}
\ee
with a single centre on the
negative $z$-axis: ${\bf x}_1\equiv (0,0,-R^2/4)$. Specifically:
\bea
K&=&-\frac{q}{2}h_1\nn
L&=&1+\frac{Q-q^2}{4}h_1\nn
M&=&\frac{3q}{4}-\frac{3qR^2}{16}h_1
\eea
with ${\bf x}={\bf x}_1$ a
coordinate singularity corresponding to the event horizon of the black
ring with topology $S^1\times S^2$. The radius of the $S^2$ is $q/2$
and that of the $S^1$ is $l$ defined by
\be
\label{defnofl}
l\equiv \sqrt{3 \left[
\frac{(Q-q^2)^2}{4q^2} - R^2 \right] } \ .
\ee
It was argued in
\cite{eemr} that demanding that this is positive ensures that the
solution is free of closed time-like curves (CTCs).

The ADM mass and angular momenta of the solution are given by
\cite{eemr}
\bea M_{ADM}&=& \frac{3\pi}{4G}Q\,,\nn
J_1&=&\frac{\pi}{8G} \,
q \, (3Q-q^2) \, , \nonumber
\\
J_2&=&\frac{\pi}{8G}\, q \,
(6R^2+3Q-q^2)\, .
\label{jpsi}
\label{adm}
\eea
The total electric charge ${\cal Q}$ satisfies $M = (\sqrt{3}/2)
{\cal Q}$, consistent with saturation of the BPS bound of
\cite{Gibbons:1993xt}.

As noted in \cite{eemr}, if we set $R=0$ then we find that the
solution becomes the black hole solution of \cite{Tseytlin:1996as,bmpv},
but we note that a new
condition needs to be imposed for the absence of CTCs:
$4Q^3>q^2(3Q-q^2)^2$. If we further set
$3Q=q^2$ then we obtain the static black hole solution.

\subsection{A generalisation} One natural generalisation of the black ring
solution of \cite{eemr} is to leave $H,K$, and $L$ unchanged and
modify $M$ as follows:
\be\label{emgen}
M=\frac{3qz}{4}-\frac{3qR^2z}{16}h_1
\ee
for constant $z$. Clearly this modification leaves $f$ unchanged.
The change in $\omega$ is easily calculated and in the
$(x,y,\phi_1,\phi_2)$ coordinates we find that $\omega_{\phi_i}\to
\omega_{\phi_i}+\delta\omega_{\phi_i}$ with $\omega_{\phi_i}$ as in
\p{omegas} and
\bea
\label{delomegas}
\delta\omega_{\phi_1}&=&\frac{3q}{4}(z-1)(1-x) \nn
\delta\omega_{\phi_2}&=&\frac{3q}{4}(z-1)(y+1) \ .
\eea
We will not analyse the global structure of these
solutions in detail here.
However, we note that $\partial_\psi$ remains space-like near
the origin (at $x=1$, $y=-1$) which is necessary
for the absence of CTCs. This would not
not have been the case for different choices of the constant term in
\p{emgen}. We also note that
$\omega_{\phi_2}=0$ at $y=-1$ and that $\omega_{\phi_1}=0$ at $x=1$.
However $\omega_{\phi_1}\ne0$ at $x=-1$, which indicates the presence
of a Dirac-Misner string whose removal seems to require periodically
identifying the time coordinate.
Finally, it is interesting to observe that ${\bf x}={\bf x}_1$
is still just a coordinate singularity for these solutions: the
analysis of \cite{eemr}, or the one we present below, goes through
straightforwardly with the only essential change being that $l$
gains a $z$ dependence via $R^2\to zR^2$ in \p{defnofl}.
Perhaps the
Euclidean version of these solutions are worth studying further.

\section{Multi-black ring solutions.}  We now come to
the main results of the paper. We keep $H=1/r$ which means that
the Gibbons-Hawking base space remains as $\bR^4$.  For $K,L$ and $M$
we consider the natural generalisation to a multi-centred
ansatz
\bea\label{mbra}
K&=&-\frac{1}{2} \sum_{i=1}^N q_i h_i
\nn
L&=&1+ \frac{1}{4} \sum_{i=1}^N (Q_i - q_i{}^2) h_i
\nn
M&=& \frac{3}{4} \sum_{i=1}^N q_i - \frac{3}{4}
\sum_{i=1}^N q_i |{\bf{x}}_i|h_i\eea with
$h_i=1/|{\bf x}-{\bf x}_i|$ and $Q_i, q_i$ are constants.
It is easy to incorporate an extra  $z_i$ constant as above, but we shall
not do so for simplicity.
We take $Q_i\ge q_i{}^2$ in order to ensure that $f\ge 0$.
The constant term in $M$ has been chosen to ensure
that $\partial_\psi$ remains spacelike at ${\bf{x}}=0$.
The solution is fully specified after solving \p{hatomeq}, which is
not simple to do in general. We will solve this equation below
when all of the poles are located on the $z$-axis, and argue that
regular solutions without CTC's exist.
It is easy, however, to determine the mass of the general solution
and we find
\be
M_{ADM}= \frac{3\pi}{4G}[ \sum_{i=1}^{N}(Q_i-q_i^2)
+ (\sum_{i=1}^{N} q_i)^2] \ .
\ee

We now analyse what happens as ${\bf x} \rightarrow {\bf x}_i$ for
some fixed $i$. To do
so we first make a rotation so that ${\bf{x}}_i$ is at
$(0,0,-R_i{}^2/4)$
and set up new spherical polar coordinates
$(\epsilon_i,\theta_i,\phi_i)$ in $\bR^3$ centred on ${\bf x}_i$ and
then consider an expansion in $\epsilon_i$.

After doing this, and solving \p{hatomeq}, we find
a coordinate singularity at $\epsilon_i=0$.  Motivated by a similar
analysis in \cite{eemr} we then introduce new coordinates
\bea\label{newcoords}
dt&=&dv+(\frac{b_2}{\epsilon_i^2} +
\frac{b_1}{\epsilon_i})d\epsilon_i
\nn
d\psi&=&d\phi_i'+2 (d\psi'+
\frac{c_1}{\epsilon_i}d\epsilon_i)
\nn
\phi_i&=&\phi_i'
\eea
for constants
$b_j$ and $c_j$.  In order to eliminate a $1/\epsilon_i$ divergence in
$g_{\epsilon_i\psi'}$ and a $1/\epsilon_i^2$ divergence in
$g_{\epsilon_i\epsilon_i}$ we take $b_2=q_i^2 l_i/8$ and
$c_1=-q_i/(2l_i)$ where
\be
l_i \equiv \sqrt{3 \left[
\frac{(Q_i-q_i^2)^2}{4q_i^2} - R_i^2 \right] },
\ee
which we take to
be positive.  A $1/\epsilon_i$ divergence in
$g_{\epsilon_i\epsilon_i}$ can be eliminated by a suitable choice for
$b_1$, whose explicit expression is not illuminating (for the single
black ring solution, $b_1 =
(2q_i{}^2 +Q_i)/(4l_i) + l_i(Q_i-q_i^2)/(3 R_i^2)$.
Note the similarity of these
constants with analogous constants appearing in \cite{eemr},
despite the fact that we are using different coordinates.)
The metric can now be
written
\bea
&& ds^2 = -\frac{256\epsilon_i^4}{q_i^4R_i^4} dv^2 -
\frac{4}{l_i} dv d\epsilon_i + \frac{32 \epsilon_i^3 \sin^2
\theta}{q_iR_i^4} dv d\phi_i'
\nn
&+& \frac{8\epsilon_i}{q_i} dv d\psi'
+ l_i^2 d{\psi'}^2 + \frac{q_i^2}{4} \left[d\theta_i^2
+ \sin^2 \theta_i d\phi_i'^2\right]
\nn
&+&2g_{\epsilon_i \phi_i'} d\epsilon_i d\phi_i'
+ 2g_{\epsilon_i\psi'}d\epsilon_i d\psi'
+g_{\epsilon_i\epsilon_i}d\epsilon_i^2
\nn
&+&2g_{\psi'\phi_i'}d\psi' d\phi_i'
+2 g_{v \theta_i} dv d \theta_i
+2 g_{\psi' \theta_i} d \psi' d \theta_i
\nn
&+& 2 g_{\epsilon_i \theta_i} d \epsilon_i d \theta_i
+ 2 g_{\theta_i \phi_i'} d \theta_i d \phi_i'
+ \ldots
\eea
where
$g_{\epsilon_i\psi'}$ and $g_{\epsilon_i\epsilon_i}$
are ${\cal O}(\epsilon_i^0)$; $g_{\psi'\phi_i'}$ and $g_{\epsilon_i
\theta_i}$ are ${\cal
O}(\epsilon_i)$; $g_{v \theta_i}$ is ${\cal O}(\epsilon_i^5)$;
$g_{\psi' \theta_i}$ is ${\cal O}(\epsilon_i^2)$; and
$g_{\theta_i \phi_i'}$ is ${\cal O}(\epsilon_i^4)$
whose explicit forms are unimportant for our considerations here, and
the ellipsis denotes terms involving sub-leading (integer) powers of
$\epsilon_i$ in all of the metric components explicitly indicated.

The determinant of this metric is analytic in $\epsilon_i$. At $
\epsilon_i=0$ it
vanishes if and only if $\sin^2 \theta_i = 0$, which just
corresponds to
coordinate singularities.
It follows that the inverse metric is also
analytic in $\epsilon_i$ and hence the above
coordinates define an analytic
extension of our solution through the surface $\epsilon_i=0$.

The supersymmetric Killing vector field $V = \partial_v$ is
null at $\epsilon_i=0$. Furthermore $V_\mu dx^\mu = -(2/l_i)
d\epsilon_i$ at
$\epsilon_i=0$, so $V$
is normal to the surface $\epsilon_i=0$. Hence $\epsilon_i=0$
is a null hypersurface and
a Killing horizon of $V$, i.e., the black
ring has an event horizon which is
the union of the Killing Horizons for each $\epsilon_i=0$.

In the near horizon limit defined by scaling $v\to v/\delta$,
$\epsilon_i\to \delta\epsilon_i$ and then taking the limit
$\delta\to 0$, we find that
the metric is locally the product of $AdS_3$ with radius $q_i$ and a
two-sphere of radius ${q_i}/2$,
in agreement with \cite{eemr}.

We can read off the geometry of a spatial cross-section of the
horizon:
\be
ds^2_\mathrm{horizon} = l_i^2 d{\psi'}^2 +\frac{q_i^2}{4}
\left(d\theta_i^2 + \sin^2{\theta_i} d\phi_i'^2 \right) \ .
\ee
We see that the horizon has geometry $S^1 \times S^2$, where the $S^1$
and round $S^2$ have radii $l_i$ and $q_i/2$, respectively. This is
precisely the geometry of the event horizon for a single
supersymmetric black ring.

The above calculation shows that as one goes near to a pole the
other poles have a sub-dominant effect.
Thus, if one of the poles lies at the origin,
i.e. we set one of the $R_i$ to zero, a similar analysis should
reveal a Killing horizon with spherical
topology corresponding to the rotating black hole solution of
\cite{Tseytlin:1996as,bmpv} (see \cite{gauntlett:99}
for a discussion of the near horizon geometry).

\subsection{Discussion}
We have just shown that the multi-ring solutions all
have Killing horizons with topology
$S^1\times S^2$. The solutions are invariant under the action
of $\partial_\psi$ (which equals $(1/2)\partial_{\psi'}$)
and the $S^1$ direction of each
horizon lies on an orbit of this vector field. To gain
some insight into the solutions, it is useful to consider
such orbits in $\bR^4$
which generically lie inside a two-plane.

In addition to the coordinates $(r,\theta,\phi,\psi)$ and
$(\rho,\Theta,\phi_1,\phi_2)$ that we have already introduced for
$\bR^4$ it will be useful to also introduce $(r_1,\phi_1,r_2,\phi_2)$,
defined by
\be
r_1=\rho\cos\Theta,\qquad r_2=\rho\sin\Theta \ .
\ee
Then $(r_i,\phi_i)$ each label an $\bR^2$ in polar coordinates
and together these give $\bR^4$. We first note that any point in $\bR^3$
labelled by $(r_0,\theta_0,\phi_0)$, $r_0\ne 0$,
defines an $S^1$, parametrised by
$\psi$, with radius $r_0$ lying in a two-plane specified
by $(\theta_0,\phi_0)$. Note also that all of these $S^1$'s are
concentric, with common centre $r=0$. For
example the point $\theta=\pi$ and
$r=R^2/4$, corresponding to the single black ring solution,
defines an $S^1$ lying just in the $(r_2,\phi_2)$ plane,
while the point $\theta=0$ and $r=R^2/4$ defines an $S^1$ lying in the
orthogonal $(r_1,\phi_1)$ two-plane, both centred at the origin.

This provides us with a natural interpretation of our
multi-black ring solutions: the
location of the pole in $\bR^3$ corresponds
to a different plane in $\bR^4$ (at asymptotic infinity) in which the
$S^1$ of the ring lies. Moreover, all of these $S^1$s are concentric.
All of the poles lying in a given direction starting from the origin
in $\bR^3$, for example the
negative $z$-axis,  correspond to
$S^1$s lying in the same two-plane.
If the poles lie on a single line passing through the
origin in $\bR^3$, for example the $z$-axis,
then the black rings lie in one of two orthogonal two-planes. This
is consistent with the fact that, as we shall see,
both $\partial_{\phi_i}$ are Killing vectors of the solution in
this case. Finally, our solutions also allow for the possibility of
a single black hole being located at
the common centre of all of these rings.

\subsection{Poles on the $z$-axis.} We now consider the solutions
with all poles located along the $z$-axis, where we can analyse the
solutions in more detail. Consider the general solution \p{mbra}
with ${\bf x_i}=(0,0,-k_iR_i^2/4)$ and $k_i=\pm 1$. Thus
\be
h_i=(r^2+\frac{k_iR_i^2}{2}r\cos\theta+\frac{R^4_i}{16})^{-1/2} \ .
\ee
We can solve \p{hatomeq} with $\hat\bomega$
only having a non-zero $\phi$ component, $\hat\omega_\phi$, that is
a function of $r$ and $\theta$ only. In particular,
these solutions have an extra $U(1)$ symmetry generated by $\partial_\phi$.
Since $\partial_\psi$ is also Killing, both $\partial_{\phi_i}$
are Killing as claimed above.

To solve \p{hatomeq} we write
$\hat\omega=\hat\omega^L+\hat\omega^Q$ where $\hat\omega^L$ is linear
in the parameters $q_i$ and independent of $Q$
and $\hat\omega^Q$ contains the dependence on $Q$.
We find
\be\label{linom}
\hat\omega^{L}=-\sum_{i=1}^{N}\frac{3q_i}{4}
[1-(r+\frac{R_i^2}{4})h_i](\cos\theta+k_i)d\phi
\ee
and
\bea
\hat\omega^{Q}&=&-\frac{3}{64}\sum_{i<j}
\frac{q_iq_j(\Lambda_j-\Lambda_i)h_ih_j}
{(k_iR_i^2-k_jR_j^2)}
\bigg[\frac{16}{h_i^{2}}+\frac{16}{h_j^{2}}
\nn
&-&\frac{32}{h_ih_j}-(k_iR^2_i-k_jR_j^2)^2\bigg] d \phi
\eea
where
\be\label{defnlam}
\Lambda_{i}\equiv \frac{(Q_i-q_i^2)}{2q_i} \ .
\ee

By considering the asymptotic form of the solution we find that
the angular momentum are given by
\bea\label{diffjays}
J_{1}&=&\frac{\pi}{8G}\bigg[ 2(\sum_{i=1}^{N} q_i)^3+3(\sum_{i=1}^{N}
q_i)\sum_{j=1}^{N}(Q_j-q_j^2)
\nn
&-&3\sum_{i=1}^{N} q_iR^2_i(k_i- 1)\bigg]\nn
J_2&=&J_1+\frac{3\pi}{4G}(\sum_{i=1}^{N}q_iR^2_ik_i) \ .
\eea

We would also like to check whether there are any Dirac-Misner strings
that might require making periodic identifications of the time coordinate.
We demand that $\omega_{\phi_1}$ vanishes at $\theta=\pi$,
which is $r_1=0$, where
$\phi_1$ is not well defined. Similarly we demand
that $\omega_{\phi_2}$ vanishes at $\theta=0$, which is $r_2=0$.
Now since $\hat\omega$ only has a $\phi$ component
for poles lying on the $z$-axis we observe that $\omega_{\phi_{1,2}}=
\omega_\psi(1\pm\cos\theta)+\hat\omega_\phi$. Thus we
demand that $\hat\omega=0$ at $\theta=0,\pi$.

The expression for $\hat\omega^{L}$ in \p{linom} satisfies
these conditions. What about $\hat\omega^{Q}$? Consider two poles
for simplicity. If $\Lambda_{1}\ne \Lambda_2$, then
analysing $\hat\omega^{Q}$ at $\theta=0,\pi$, we find that it
is constant for values of $z$ between the poles and vanishes otherwise.
One might try to remove the constant by introducing another
coordinate patch to cover this region with a shift in
the time coordinate by an appropriate constant multiple of $\phi_i$.
However, since $\phi_i$ are periodic coordinates,
this means that the time coordinate must be periodically identified.
Thus we conclude that to avoid CTCs $\Lambda_{1}=\Lambda_2$.
It would be interesting to obtain a physical understanding of
this constraint, and its obvious generalisation to arbitrary numbers
of rings.

After imposing $\Lambda\equiv\Lambda_1=\Lambda_2=\dots$, the
radii of the rings is given by
\be
l_i \equiv \sqrt{3 \left[\Lambda^2 - R_i^2 \right] } \ .
\ee
We conclude that we can place the two rings anywhere on the $z$-axis
provided that $R_i^2<\Lambda^2$. It is also interesting to
observe that as the location of the pole goes to larger
values of $|z|$, i.e. as $R_i$ increases, the
circumference of the rings get uniformly smaller, perhaps
contrary to one's intuition.

To ensure that there are configurations with no CTC's we also demand that
the determinant of the $(\phi,\psi)$ part of the five-dimensional
metric remains positive. We have analysed this numerically and found
that this can be achieved for specific values of the parameters.
We leave more detailed analysis for future work.

It is interesting to ask whether there are configurations of
two (say) black rings with the same charges as the black hole solution
of \cite{Tseytlin:1996as,bmpv}. Since the black hole has
$J_1=J_2$, we conclude from \p{diffjays} that
we need to locate one pole on the positive axis and one on
the negative axis, $k_2=-k_1$, and in addition set $q_1R_1^2=q_2R_2^2$.
We denote the black hole parameters by $\bar Q, \bar q$ and define
$j\equiv \bar q(3\bar Q-\bar q^2)/2$, with  $\bar Q>0$.
By matching the mass
and the angular momentum of the black hole with the two rings
we deduce that
\bea
\bar Q&=&2\Lambda(q_1+q_2)+(q_1+q_2)^2\nn
j&=&(q_1+q_2)^3+3\Lambda(q_1+q_2)^2+3(q_1 R_1^2) \ .
\eea
We demand that $\Lambda^2>R_i^2$. The area of the black hole
event horizon is given by
\be
A_{BH}=2\pi^2\sqrt{\bar Q^3-j^2}
\ee
and we demand that $j^2<\bar Q^3$, which
ensures that the black hole has no CTCs, and is equivalent to
\bea
8\Lambda^3(q_1+q_2)^3+3\Lambda^2(q_1+q_2)^4&>&9q_1^2R_1^4
\nn
+6q_1R_1^2(q_1+q_2)^3&+&18\Lambda q_1R_1^2(q_1+q_2)^2 \ . \nn
\eea
This can be satisfied if we choose small enough $q_1R_1^2$.
Observe that if we set $q_1=q_2$, then we can solve for $q_1$ so that
$j=0$:
\be
q_1=\frac{-3\Lambda+\sqrt{9\Lambda^2-6R_1^2}}{4} \ .
\ee
However, we find that the solution has $\Lambda$ and $q_1$ having
opposite signs, but they should have the same sign from \p{defnlam}.

It is interesting to compare the black hole area
with the sum of the areas of the horizons of the two black rings:
\bea
A_{Rings}&=&2\pi^2(q_1^2l_1+q_2^2l_2)\nn
&=&2\pi^2{\sqrt 3}\left( q_1^2\sqrt{\Lambda^2-R_1^2}
+q_2^2\sqrt{\Lambda^2-R_2^2}\right) \ .
\nn
\eea
It is remarkable that parameters can be chosen so that these
areas are equal. For example, if we take
$R_1=R_2$ and
\bea
q_1&=&q_2=\frac{1}{18(\Lambda^2-R_2^2)} \bigg[
-16\Lambda^3+18R_2^2\Lambda
\nn
&+&\sqrt{256\Lambda^6-576\Lambda^4R_2^2
+405R_2^4\Lambda^2-81R_2^6} \bigg]
\eea
Furthermore (again with $q_1=q_2$, $R_1=R_2$),
 we can also arrange for $A_{Rings}<  A_{BH}$;
 for example $A_{Rings}=\frac{1}{2}  A_{BH}$
can be achieved by setting
\be
q_2 = \frac{9 R_2^4}{8 \Lambda (8 \Lambda^2 -9 R_2^2)} \ .
\ee
Lastly, we can also arrange for $A_{Rings}> A_{BH}$;
for example $A_{Rings}=2 A_{BH}$ with
$R_1 = R_2$ and $q_1 = q_2$ can be achieved with
\bea
q_2 &=& \frac{1} {45 (\Lambda^2-R_2^2)} \bigg[
36 \Lambda R_2^2 -32 \Lambda^3
\nn
&+& \sqrt{1024 \Lambda^6-2304 \Lambda^4 R_2^2+1701 \Lambda^2 R_2^4
-405 R_2^6} \bigg]
\nn
\eea

For the specific values
\be
\Lambda=3/2,\qquad R_2=1
\ee
we obtain $q_1=\frac{\sqrt{41}-6}{5}$ for $A_{Rings}= A_{BH}$,
$q_1=\frac{1}{12}$ for $A_{Rings}=\frac{1}{2} A_{BH}$
and $q_1 = \frac{2}{25}$ for $A_{Rings}=2 A_{BH}$.
In all three cases, $\Lambda^2>R_i^2$ and  $j^2<\bar Q^3$.

A numerical investigation of these solutions suggests that they
do not possess closed timelike curves. It would be  interesting to
verify this result analytically.

\section{Multi-black strings} In the infinite radius limit the black
ring solution of \cite{eemr} gives rise to the black string
solution of \cite{bena}. This solution can easily be constructed in
Gibbons-Hawking form with an $\bR^4$ base with $H=1$, $\bchi=0$ and
\be K=-\frac{q}{2r},\quad L=1+\frac{Q}{r},\quad M=-\frac{3q}{4r} \ee
The expression for $\omega$ is given by
\bea
\omega=-(\frac{3q}{2r}+\frac{3qQ}{4r^2}+\frac{q^3}{8r^3})d\psi \ .
\eea
Following \cite{bena}, we choose $Q^2\ge q^2$ to eliminate CTCs at r=0.

We note that modifying $M\to
-3qz/4r$ leads to $\omega\to\omega +\delta\omega$ with
\be
\delta\omega=\frac{3q(1-z)}{4}(\frac{1}{r}d\psi+
\cos\theta d\phi)
\ee
which has Dirac-Misner strings which seem to require periodic
identification of the time coordinate.

A more physical multi-string generalisation, with all $z=1$, is given by
$H=1$ and
\bea
K&=&-\frac{1}{2}\sum_{i=1}^{N} q_ih_i
\nn
L&=&1+\sum_{i=1}^{N} Q_i h_i
\nn
M&=&-\frac{3}{4}\sum_{i=1}^{N} q_i h_i
\ee
with $h_i=1/|{\bf x}-{\bf x}_i|$. The
solution is fully specified after solving \p{hatomeq}, which can be done
explicitly for special cases.

\section{Final comments}
Using the techniques of \cite{gghpr}, we have
found several supersymmetric
generalisations of the supersymmetric black ring of \cite{eemr}
and the supersymmetric black string of \cite{bena}.
The most interesting construction describes concentric black rings
with a possible black hole at the origin.
We found configurations of two black rings with the same asymptotic
charges as a rotating black hole but with entropy greater than, equal to,
or less than that of the black hole. We hope to improve upon our
numerical investigations and prove that these black rings are
indeed free of CTCs. Accounting for the entropy of the black ring
solutions found here and in \cite{eemr} is an important outstanding issue.

Since multi-supersymmetric black holes are known
to exist \cite{gauntlett:99},
it is natural to ask if the solutions presented here can be generalised
to give solutions with black rings centred around each of the black holes.
While this may be possible, more
elaborate tools will be required to find them.
The reason is simply that the multi-black hole solutions themselves cannot
be constructed using the Gibbons-Hawking ansatz that
we have used here, which
assumed that the solution is invariant under the
tri-holomorphic Killing vector field $\partial_\psi$.
To see this recall that the multi-black hole solutions
have $\bR^4$ for the base manifold $M_4$ in \p{metric} and
$f^{-1}$ a harmonic function
on $\bR^4$ with isolated point singularities. However,
after writing $\bR^4$ in Gibbons-Hawking form \p{ghflat}, we find that
it is only the single black hole solution that is invariant
under $\partial_\psi$.


\begin{acknowledgments}
We thank Fay Dowker for very helpful discussions. J.B.G.
thanks EPSRC for support.
\end{acknowledgments}


\end{document}